\documentclass[12pt]{article}

\usepackage{amsmath,amsfonts,amssymb}
\usepackage{graphicx}
\usepackage{psfrag}
\usepackage{enumerate}
\usepackage{mathrsfs}
\newcommand\blfootnote[1]{%
  \begingroup
  \renewcommand\thefootnote{}\footnote{\hspace{-6mm}#1}%
  \addtocounter{footnote}{-1}%
  \endgroup
}

\newcommand{\be}{\begin{equation}}
\newcommand{\ee}{\end{equation}}

\newcommand{\eeq}{\end{eqnarray}}

\def\[{\left [}
\def\]{\right ]}
\def\({\left (}
\def\){\right )}

\def\r2{\sqrt{2}}

\newcommand{\bbibitem}[1]{\bibitem{#1}\marginpar{#1}}

\def\Label#1{\label{#1}%
  \smash{\hbox to0pt{\raise1ex\hbox{\tiny[#1]}\hss}}}
\def\noLabels{\let\Label=\label}
\def\nobbibitem{\let\bbibitem=\bibitem}

\begin{document}
\noLabels % uncomment for final production
\nobbibitem % uncomment for final production
\clearpage\thispagestyle{empty}

\begin{center}

%\begin{flushright} \vspace{-3cm}
%{\small UPR-1222-T}  \\
%\end{flushright}
%\vspace{1cm}

{\Large \bf  A hole-ographic spacetime}

\vspace{7mm}

Vijay Balasubramanian$^{a,b,c}$, Borun D. Chowdhury$^{d,e}$, Bart{\l}omiej Czech$^{d,f}$, \\
Jan de Boer$^{d}$, Michal P. Heller$^{d,g}$

\blfootnote{\tt vijay@physics.upenn.edu,czech@stanford.edu,bdchowdh@asu.edu,m.p.heller,J.deBoer@uva.nl}
%\\

%\vspace{5mm}

%\bigskip\centerline{$^a$\it Department of Physics and
%Astronomy}
\bigskip\centerline{$^a$\it David Rittenhouse Laboratories, University of Pennsylvania}
\smallskip\centerline{\it 209 S 33$^{\rm rd}$ Street, Philadelphia, PA 19104, USA}
%\bigskip\medskip
\bigskip\centerline{$^b$\it Laboratoire de Physics Th\'{e}orique, \'{E}cole Normale Sup\'{e}rieure}
\smallskip\centerline{\it 24 rue Lhomond, 75005 Paris, France}
\bigskip\centerline{$^c$\it CUNY Graduate Center, Initiative for the Theoretical Sciences}
\smallskip\centerline{\it 365 Fifth Avenue, New York, NY 10016, USA}
%\bigskip\medskip
\bigskip\centerline{$^d$\it Institute for Theoretical Physics, University of Amsterdam}
\smallskip\centerline{\it Science Park 904, Postbus 94485, 1090 GL Amsterdam, The Netherlands}
%\bigskip\medskip
\bigskip\centerline{$^e$\it Department of Physics, Arizona State University}
\smallskip\centerline{\it Tempe, Arizona 85287, USA}
%\bigskip\medskip
\bigskip\centerline{$^f$\it Department of Physics, Stanford University}
\smallskip\centerline{\it 382 Via Pueblo Mall, Stanford, CA 94305-4060, USA}
%\bigskip\medskip
\bigskip\centerline{$^g$\it National Centre for Nuclear Research}
\smallskip\centerline{\it Ho{\.z}a 69, 00-681 Warsaw, Poland}
%\vfil
\end{center}

\vspace{5mm}

\begin{abstract}
\noindent
We embed spherical Rindler space -- a geometry with a spherical hole in its center -- in asymptotically AdS spacetime and show that  it carries a  gravitational entropy proportional to the area of the hole. Spherical AdS-Rindler space is holographically dual to an ultraviolet sector of the boundary field theory given by restriction to a strip of finite duration in time.       Because measurements have finite durations,  local observers in the field theory can only access information about bounded spatial regions.  We propose a notion of differential entropy that captures  uncertainty about the state of a system left by the collection of local, finite-time observables.   For two-dimensional conformal field theories we use holography and the strong  subadditivity of entanglement to propose a formula for differential entropy and show that it precisely reproduces the areas of  circular holes in AdS$_{3}$.   Extending the notion to field theories on strips with variable durations in time, we show more generally that  differential entropy computes the areas of all closed, inhomogenous curves on a spatial slice of AdS$_3$.   We discuss the extension to higher dimensional field theories, the relation of differential entropy to entanglement between scales, and some implications for the emergence of space from the RG flow of entangled field theories.
\end{abstract}

\setcounter{footnote}{0}
\newpage
\clearpage
\setcounter{page}{1}

%\tableofcontents
%\section{Introduction}
%\label{intro}

\section{Introduction}

We  recently reported a calculation of the gravitational entropy of spherical Rindler space \cite{lastpaper}.\footnote{Recall that acceleration horizons can carry gravitational entropy, similarly to black hole and cosmological horizons \cite{laflamme}.} This is a region of Minkowski space, which consists of  points that can exchange signals with at least one out of a family of radially accelerating observers. The salient feature of spherical Rindler space is that it has a hole in its center, and the area of the hole in units of $4G$ measures the gravitational entropy.

The present paper embeds this construction in anti-de Sitter space, which allows us to study  a spacetime with a hole holographically. In analogy to spherical Rindler-Minkowski space, we consider a family of observers who accelerate away from the center of anti-de Sitter (AdS) space. The worldlines of these observers are causally disconnected from a spherical hole in AdS, whose coordinate radius we call $R_0$.    In the holographic description of AdS space, the exterior of the hole ($R > R_0$) and the interior of the hole ($R < R_0$) should be associated to the ultraviolet (UV) and, respectively, the infrared (IR) of the dual field theory, a qualitative fact that the holographic renormalization group attempts to capture \cite{holoRG1, holoRG2, holoRG3}.  Therefore, given the relation between geometry and entanglement discovered in \cite{rt} and studied in \cite{swingle1, mav1, mav2, swingle2, rqg, myers1, myers2, maulik},\footnote{See also \cite{kl} for an early qualitative formulation and \cite{recon1, recon2, recon3, recon4} for procedures to reconstruct the spacetime from entanglement.} it would be natural to guess that areas of radial surfaces in AdS space are related to some sort of UV/IR entanglement in the dual field theory \cite{uvir}.   But what is the appropriate division into UV and IR observables? 

The relevant separation is in terms of the timescales over which local observers in the field theory can make measurements.   To see this simply, project future- and past-directed outward light rays from the edge of the hole, $R=R_0$. These light rays reach the boundary in finite global time $\pm T_0$. Given the structure of holographic duality, this means that  information about the interior of the hole can only be locally encoded in the boundary field theory  {\it outside} the time interval $-T_0 < T < T_0$.  Equivalently, local observables in the boundary field theory {\it within} a strip of finite duration $2T_0$ fully encode the physics in the $R>R_0$ region of anti-de Sitter space.

Restricting attention to observables over a finite duration in a field theory may seem exotic.   It is more common to restrict oneself to observables in a  finite region of space.  But whenever we specify initial data on a bounded spatial region $D$ on a time slice, we implicitly specify it everywhere in the domain of dependence of $D$, i.e. in the spacetime region whose physics is entirely determined by the data on $D$.\footnote{See \cite{Casini:2006es} for how this observation relates strong subadditivity to the $c$-theorem in two dimensions.} Thus, it is more precise to say that we understand well how to restrict to the observables of a field theory in a domain of dependence of a spatial region (a causal diamond in the case when the spatial region is an interval).    As we will see below, the way to restrict a field theory to local observables of a finite duration in time is to assemble the associated  space from the causal diamonds of local observers.  Specifically, a strip of time duration $2T_0$ can be regarded as the union of causal diamonds of intervals of length $2T_0$, where we associate each causal diamond to a local observer.   This union is a holographic representation of the way in which we construct spherical-Rindler-AdS space as the union of a set of regions that can exchange signals with a single accelerated observer.   Each of these bulk regions will be  related to a single causal diamond on the boundary. 

When we specify initial data on every spatial interval of length $2T_0$, does this not specify the entire future development of the system?  It does not -- the missing data is the pattern of entanglement. The simplest example is two spins: all states of the form
\begin{equation}
|\Psi\rangle = \cos\phi\, |\!\uparrow\uparrow\rangle 
+ e^{i \theta} \sin\phi \,|\!\downarrow\downarrow\rangle
\end{equation}
restrict to the same density matrices of the individual spins, independent of $\theta$.     In general, it is difficult to quantify the uncertainty in a quantum state that remains after restricting  to finite time local observables, or equivalently to the information that can be retrieved from a set of finite duration causal diamonds (but see \cite{hartle}). Here we use holography to propose  a formula (eq.~\ref{contobs}) quantifying this ``differential entropy'' in the vacuum of two-dimensional field theory. It is a combination of entanglement entropies or, equivalently, causal holographic information \cite{CHI} of spatial regions of size $2T_0$. Formula~(\ref{contobs}) precisely reproduces the areas of circular holes in AdS$_3$ and saturates the strong subadditivity bound for quantum information.   In other words, we show that a circular hole in AdS$_3$ has maximal area that is consistent with strong subadditivity in the dual field theory.   Generalizing these results, we propose a measure of differential entropy for two-dimensional field theories on a strip of variable duration in time (eq.~\ref{genbound}) and show that this quantity exactly reproduces the lengths of all inhomogenous closed curves on a spatial slice in AdS$_3$.    As we will discuss,
this novel quantity does not appear to have come from a reduced density matrix, so it should not be interpreted as entanglement entropy in a strict sense. This observation suggests that the Hilbert space of quantum gravity does not factorize between the inside and outside of a closed surface, contradicting the expectation from the presumed locality of spacetime.    Thus, our results raise questions about the validity of local low energy effective field theory in a theory containing gravity. This issue does not affect typical low energy experiments such as scattering problems, but it is relevant to discussions of holographic entanglement entropy.

The organization of this paper is as follows. In Sec.~\ref{coords} we introduce the spherical-Rindler-AdS space and discuss its field theory dual, which is a finite time strip of the field theory living on the global boundary. We discuss the assembly of this sector of the field theory  from causal diamonds, which correspond to individual accelerating observers in the bulk. In Sec.~\ref{thebound} we focus on two-dimensional field theory and propose a measure of differential entropy, quantifying ignorance of the underlying quantum state that remains after making all local finite-time observations.   We show that the resulting quantity reproduces the areas of spatially inhomogenous holes in AdS$_3$ and saturates the strong subadditivity bound for quantum information.  The paper closes in Sec.~4 with a discussion.

\section{Spherical Rindler-AdS space}
\label{coords}

\subsection{A spherical hole in Minkowski space -- a review}
We start with a lightning review of the spherical Rindler space cut out of flat spacetime \cite{lastpaper}.  In radial coordinates,
\begin{equation}
ds^2 = -dT^2 + dR^2 + R^2 d\Omega_{d-1}^2 \, ,
\end{equation}
trajectories with a constant radial acceleration take the form
\begin{equation}
T(t) = r \sinh t \qquad {\rm and} \qquad R(t) = R_0 + r \cosh t\,,
\label{rindlerrules}
\end{equation}
with proper acceleration $a = r^{-1}$ and proper time along the trajectory given by $r t$. To Rindlerize, we treat eqs.~(\ref{rindlerrules}) as a coordinate transformation. The Rindler coordinate $t$ parameterizes time along the accelerated trajectories while the Rindler coordinate $r$ parameterizes the acceleration. The resulting metric is:
\begin{equation}
ds^2 = -r^2 dt^2 + dr^2 + (R_0 + r \cosh t)^2 d\Omega_{d-1}^2.
\label{sphRindlMink}
\end{equation}
The quantity $R_0$ is a constant, which determines the size of the region in the center that remains out of causal contact with the accelerated observers. Said differently, $R_0$ is the radius of the hole. 

Metric (\ref{sphRindlMink}) has a horizon at $r=0$. In the parent Minkowski space, this horizon is the edge of the causal past and future of the hole. The gravitational entropy of the space (\ref{sphRindlMink}) is related to the area of this horizon in the usual way:
\begin{equation}
S = \frac{\mathcal{A}}{4G} = \frac{{\rm vol}(S^{d-1})\,R_0^{d-1}}{4G}. \label{entropyexpl}
\end{equation}
Although metric~(\ref{sphRindlMink}) is time-dependent, its near-horizon limit 
\begin{equation}
ds^2 = -r^2 dt^2 + dr^2 + R_0^2\, d\Omega_{d-1}^2
\end{equation}
is static, so one expects a well-defined and time-independent entropy. More rigorously, eq.~(\ref{entropyexpl}) can be derived using the replica trick, which generalizes the conical deficit method to rotationally non-invariant spacetimes \cite{fursaevsold, BTZconic, simonreview}.  Consider the Euclidean continuation of (\ref{sphRindlMink}):
\begin{equation}
ds^2 = r^2 dt_E^2 + dr^2 + (R_0 + r \cos t_E)^2 d\Omega_{d-1}^2.
\label{EsphRindlMink}
\end{equation}
The Euclidean time $t_E$ is an angular coordinate, whose periodicity $\beta_0 = 2\pi$ can be read off both from the regularity at the origin and from the single-valuedness of $g_{\Omega\Omega}$. Call its action $Z_1$. The action of an integer cover of (\ref{EsphRindlMink}) with periodicity $\beta = n\beta_0$ is given by:
\begin{equation}
Z_n = \frac{\beta}{\beta_0} Z_1 + (\beta-\beta_0) \frac{\mathcal{A}}{8 \pi G}.
\label{nfoldcover}
\end{equation}
The first term arises from the region away from $r=0$ while the second term, proportional to the transversal size of the locus $r=0$, is the conical surplus term. The action is related to the free energy $F(\beta)$ via $Z = \beta F(\beta)$, which means that the entropy is given by $S = (\beta \partial_\beta - 1) Z$ evaluated at $\beta_0$. Applying this to the obvious analytic continuation of (\ref{nfoldcover}) gives result (\ref{entropyexpl}).

A shortcut way to read off the entropy is to neglect the time-dependent physics away from the horizon by dropping the term $r\cosh t$ in $g_{\Omega\Omega}$. This reflects the intuition that a static horizon captures an equilibrium between a gravitational system and a heat bath so that, as a consequence, whenever two spacetimes share the same near-horizon geometry, they must also have the same entropy. For the metric (\ref{sphRindlMink}), the near-horizon limit $r\to 0$ is the same as the near-horizon limit of a Schwarzschild black hole, whose entropy is known to be (\ref{entropyexpl}). 

Both ways of deriving (\ref{entropyexpl}) are discussed in greater detail in \cite{lastpaper}. 

\subsection{Spherical Rindler-AdS space}
We start with AdS space in the global coordinates:
\begin{equation}
ds^2 = - \left( 1 + \frac{R^2}{L^2}\right) dT^2 + \left( 1 + \frac{R^2}{L^2}\right)^{-1} dR^2 + R^2 d\Omega_{d-1}^2.
\label{adsglobal}
\end{equation}
Our goal is to cut a hole in it, thereby generalizing metric (\ref{sphRindlMink}) to anti-de Sitter space. To proceed as before, we must find the radially accelerated trajectories and treat the acceleration and time along the trajectory as a pair of coordinates. The trajectories are derived in Appendix~\ref{radials}. To reach the boundary, the acceleration $a$ must exceed $L^{-1}$, so we use the parameterization $aL = \cosh\rho$. The final form of the trajectories is:
\begin{eqnarray}
R(t, \rho) & = &
\frac{L}{\sinh{\phi} \sinh\rho}\, \big(\cosh\rho 
+ \cosh\phi\cosh t\big) \label{defR}\\
T(t, \rho) & = & L \cot^{-1} 
\frac{\cosh\phi \cosh\rho + \cosh t }{\sinh\phi\, \sinh t} \label{defT}
\end{eqnarray}
Here $\phi$ is a parameter, which controls the global time at which the trajectory reaches the asymptotic boundary $R \to \infty$:
\begin{equation}
T_0 = L \tan^{-1} \sinh\phi.
\end{equation}
In other words, all trajectories with the same $\phi$ asymptote to a common radial outgoing light ray, which intersects the slice of time symmetry $T=0$ at a coordinate radius $R_0$:
\begin{equation}
T_0 = \int_{R_0}^\infty \frac{dR}{1+\frac{R^2}{L^2}} = L \cot^{-1} \frac{R_0}{L}
\qquad \Rightarrow \qquad 
R_0 = \frac{L}{\sinh\phi}.
\label{defT0}
\end{equation}
Treating eqs.~(\ref{defR}-\ref{defT}) as a coordinate redefinition, the AdS metric takes the form:
\begin{equation}
ds^2 = \frac{L^2}{\sinh^2\!\rho} \left(-dt^2 + d\rho^2 + \left(\frac{\cosh\rho
+ \cosh\phi\cosh t}{\sinh\phi}\right)^2 d\Omega_{d-1}^2 \right).
\label{sphRindlAdS}
\end{equation}
When $\rho$ ranges over positive numbers, this metric covers only the region which is causally disconnected from a sphere of radius $R_0 = L/\sinh\phi$ at $T=0$. This region is the spherical Rindler-AdS space. It contains a horizon at $\rho = \infty$, which is the limit where the accelerated trajectory becomes arbitrarily close to the light ray projected from $R=R_0$.

To highlight the analogy with metric (\ref{sphRindlMink}), perform one final change of coordinates:
\begin{equation}
\frac{d\rho}{\sinh\rho} = - dr \qquad \Rightarrow \qquad \sinh r = \frac{1}{\sinh \rho} 
= \frac{1}{\sqrt{(aL)^2-1}}.
\end{equation}
This is an AdS analogue of the definition of the Minkowski-Rindler radial coordinate $r = a^{-1}$. Now the spherical Rindler-AdS metric becomes:
\begin{equation}
ds^2 = L^2 \big(-\!\sinh^2 r \,dt^2 + dr^2\big) + 
\left(R_0 \cosh r + \sqrt{R_0^2 + L^2} \,\sinh r \cosh t \right)^2 d\Omega_{d-1}^2.
\label{diranalogue}
\end{equation}

This metric should be compared with the spherical Rindler metric (\ref{sphRindlMink}) obtained from flat space. The horizon is again at $r = 0$, which is the limit of large acceleration $a$. In the neighborhood of $r = 0$, the Euclidean continuation of (\ref{diranalogue}) looks once more like a plane in polar coordinates times a transversal sphere of constant size. This means that the computations of the gravitational entropy carried out in \cite{lastpaper} apply to metric (\ref{diranalogue}) without modifications. The result is again: 
\begin{equation}
S = \frac{\mathcal{A}}{4G} = \frac{{\rm vol}(S^{d-1})\,R_0^{d-1}}{4G}.
\end{equation}

\subsection{Field theory on a finite time interval}
\label{finitetime}

The horizons which bound the spherical Rindler-AdS space in the bulk extend to the global asymptotic boundary. They reach it at $\pm T_0$ given in eq.~(\ref{defT0}), which is the global time at which light rays projected from $R=R_0$ arrive at the boundary.  Holographically, the physics in the spherical Rindler-AdS space (the region that is causally disconnected from the interior of an $R_0$-sized hole at $T=0$) is fully encoded in local field theory observables in the time interval $(-T_0, T_0)$. 

To understand this sector, consider the way in which spherical Rindler space is constructed. It is the union of ordinary Rindler spaces for a family of radially accelerating observers. The trajectory of each of these observers starts on the boundary at time $T = -T_0$ and ends on the boundary at time $T = T_0$. Causality alone then suggests that the observations carried out by this accelerated observer must be encoded on the field theory side in the causal diamond extending between $T =\pm T_0$. Indeed, the bulk causal wedge associated with the boundary causal diamond extending between $T =\pm T_0$ is precisely the ordinary AdS-Rindler space associated with a single accelerating observer \cite{casinihuertamyers, CHI, rqg, maulik} 

\begin{figure}[t]
\centering
\includegraphics[width=.7\textwidth]{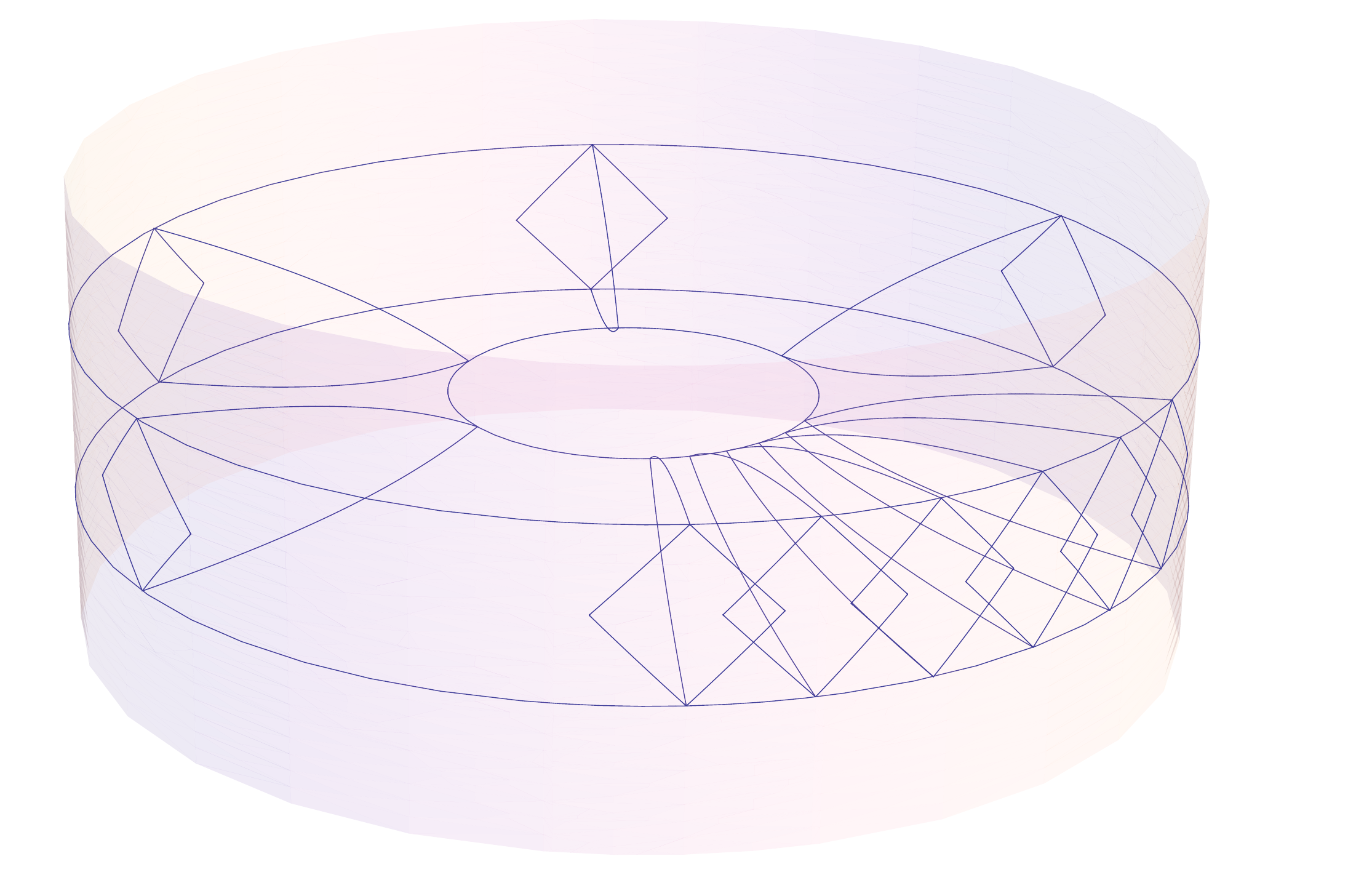}
\caption{Spherical Rindler-AdS space, with a hole inside it, is built up of regions visible to individual accelerating observers, each of whom observes physics that is holographically dual to the content of a single causal diamond. The union of the diamonds makes up a finite time strip in the boundary field theory.}
\label{CFTUV}
\end{figure}

Fig.~\ref{CFTUV} shows  spherical Rindler space as the union of  causal wedges, each of which is associated with a boundary causal diamond spanning the time interval between $T_0$ and $-T_0$. This is a pictorial representation of the definition of the spherical Rindler-AdS space, which is the union of regions that can exchange signals with at least one of a family of radially accelerated observers. The union over all the accelerating observers in the bulk corresponds to the union of all the causal diamonds on the boundary. This union is simply the strip of the boundary field theory that spans the time interval between $-T_0$ and $T_0$.

Local observers in the field theory with a finite time duration cannot access the physics of wavelengths longer that $2T_0$, because those would not fit inside any one causal diamond.   Thus, local observers can only access the physics of short wavelengths which, according to the AdS/CFT dictionary, encode bulk regions at large radii.   In this way, local observables on the strip effectively isolate a UV sector of the field theory. The construction is summarized in Fig.~\ref{CFTUV}.     

It is well understood how to restrict observations to one component of a Hilbert space that enjoys a tensor product decomposition $A \otimes B$.   In this circumstance,  the restricted observations will effectively be carried out in a density matrix obtained by tracing out the unobserved component.  The von Neumann entropy of this density matrix quantifies the entanglement between the observed and un-observed parts of the Hilbert space. The conventional example of this scenario arises in quantum field theory when we restrict observations to a spatial box.  Another example, in perturbative field theory, arises by separating the theory into UV and IR components with a momentum cutoff \cite{uvir}.   Here we are considering a subspace of observables that do not necessarily define a tensor factor of the Hilbert space.    Nevertheless, because the set of observables is incomplete, it leaves uncertainty about the underlying quantum state, e.g. about entanglement between distant causal diamonds.   We seek a measure of this uncertainty, which we will call differential entropy. A natural definition of differential entropy could be that it is equal to the maximal entropy $S_{\rm max}$ attainable by a density matrix, which correctly describes all measurements of finite-time local observers.\footnote{This is similar to the proposed field theory interpretation of causal holographic information \cite{wall}.} It is in general difficult to compute such an entropy exactly.   In the next section we propose an explicit and tractable quantity for differential entropy in two-dimensional holographic theories. The evidence for our proposal is that it reproduces the lengths of arbitrary closed curves on a time slice of AdS${}_3$.

\section{Differential entropy for 2d holographic field theories}
\label{thebound}

The gravitational entropy of spherical Rindler-AdS$_3$ space is:
\begin{equation}
S_{\rm gr} = \frac{2\pi R_0}{4G} \, .
\label{holeentropy}
\end{equation}
We would like to recover this formula from information-theoretic quantities in the dual conformal field theory.

\subsection{A formula for differential entropy}
\label{formula}

\begin{figure}[t]
\centering
\includegraphics[width=\textwidth]{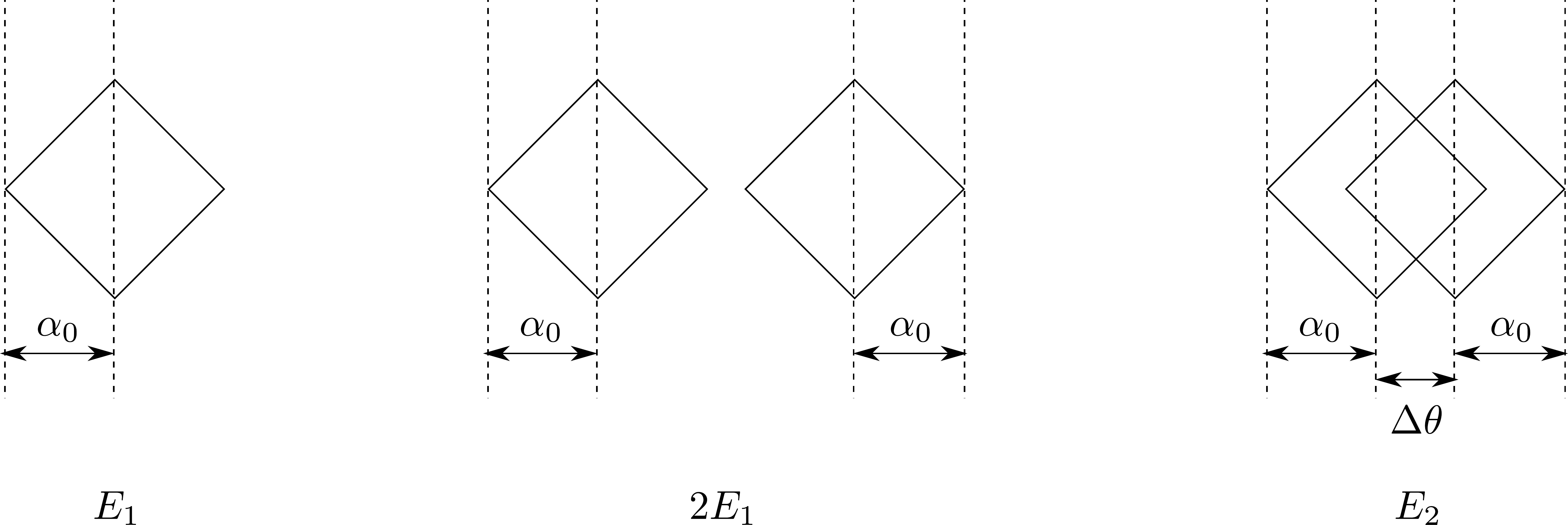}
\caption{Combinations of boundary causal diamonds considered in the derivation in Sec.~\ref{formula}.}
\label{derive}
\end{figure}

We are interested in quantifying the collective ignorance of a family of  local observers in the conformal field theory who make measurements over a finite duration $2T_0$.   First consider one observer, whose associated causal diamond is illustrated on the left of Fig.~\ref{derive}.  Measurements performed by this observer can access data on an interval of length $2T_0$ on the $T=0$ surface of the time strip.   Thus, they will effectively be carried out in a density matrix obtained by tracing out the exterior of the interval in the vacuum of the field theory.    The associated entanglement entropy, quantifying this observer's ignorance of the full state of the system, is given by
\be
S({\alpha_0}) = \frac{c}{3} \log \left( \frac{2L}{\mu} \sin\alpha_0 \right),
\label{oneobserver}
\ee
where $2\alpha_0 = 2T_0/L$ is the angular size of the interval.  The quantity $c$ is the central charge and $\mu$ is a UV cutoff.  Removing the cutoff would cause the entanglement entropy to diverge because of the large number of UV modes that straddle any spatial boundary.    By contrast, the gravitational entropy in (\ref{holeentropy}) has no UV divergence. Canceling off the UV divergence will be a guide to the correct formula for differential entropy.

If we now consider adding a second observer whose causal diamond is disjoint from the first, we could add the entanglement entropies of both observers (the center panel of Fig.~\ref{derive}). Note that this would effectively double the UV divergence of the resulting quantity.  However, our task is to consider the family of all local, finite time observers whose causal diamonds overlap. Consider two such neighboring  observers (the right panel of Fig.~\ref{derive}) with an angular separation $\Delta \theta$.   Adding their entanglement entropies clearly overcounts their ignorance of the underlying entanglement.   We might attempt to correct this overcount by subtracting the entanglement entropy of the overlap of the two causal diamonds:
\begin{equation}
E_2 = S({\alpha_0}) + S(\alpha_0) - S(\alpha_0 - \Delta\theta/2)
\label{twoobserver}
\ee
Note that the subtraction has removed some of the undesirable UV divergence: after the subtraction (\ref{twoobserver}) has the same UV divergence as (\ref{oneobserver}). 

We now consider a family of $2K$ evenly spaced observers.  Nearest neighbors have an  angular separation $\Delta\theta = \pi/K$.  Iterating (\ref{twoobserver}) gives the formula:
\be
E_{2K} = 2K \big(S(\alpha_0) - S(\alpha_0 - \Delta\theta/2) \big)
=
 \frac{2 K c}{3} \left( \log \frac{2L}{\mu} \sin\alpha_0 - 
\log \frac{2L}{\mu} \sin \Big(\alpha_0 - \frac{\pi}{2K}\Big) \right) .
\label{Kobservers}
\ee
The UV divergences cancel telescopically around the circle supporting the field theory.
The continuum limit, which gives our proposed definition of the differential entropy, is:
\be
E = \pi {d S(\alpha) \over d\alpha}\Big{|}_{\alpha_0} = {\pi c \over 3} \cot\alpha_0
\label{contobs}
\ee
Using the holographic relation $c = 3L/2G$ and the relation between the size of the hole and the duration of the time strip (\ref{defT0}) gives
\be
E = {2\pi R_0 \over 4G} = S_{\rm gr}\, ,
\ee
which precisely reproduces the area of the hole.

\paragraph{Holographic derivation: }  In a general field theory, it would be quite surprising if subtracting the entanglement entropy of overlaps was an adequate way to deal with overlapping causal diamonds.  But we are here considering theories with a holographic dual.  A simple geometric argument in such theories rationalizes our proposal.  Recall first that Ryu and Takayanagai have shown that the entanglement entropy of an interval in the field theory (\ref{oneobserver}) is equal to the length of a spatial geodesic in AdS$_3$ that subtends the boundary interval \cite{rt}.  In our case these geodesics are given by 
\be
\tan^2\tilde\theta(R) = \frac{{R^2 \tan^2\alpha_0- L^2}}{R^2+L^2}\,,
\label{rtcurve}
\end{equation}
where the geodesic is parametrized by its angular coordinate $\tilde\theta$ at each radial position $R$ in the metric (\ref{adsglobal}) and $2\alpha_0$ is the angle subtended at the boundary. Importantly, the minimal radius reached by the geodesic is
\begin{equation}
L \cot\alpha_0 = R_0.
\label{minradius}
\end{equation}
In assembling a finite time strip of the field theory from causal diamonds, we have defined $\pm T_0$ as the time at which a light ray projected from $R=R_0$ reaches the boundary and then translated it into an angular interval of size $2\alpha_0 = 2T_0 / L$. Eq.~(\ref{minradius}) states that the spatial geodesic encoding the entanglement entropy of this interval is tangent to the bulk circle we started with.  This reflects a relation between entanglement entropy and causal holographic information that holds in the vacuum of two dimensional holographic theories \cite{CHI}, but not in general settings such as excited states in higher dimensions.

\begin{figure}[t!]
\centering
\begin{tabular}{ccc}
\includegraphics[width=.34\textwidth]{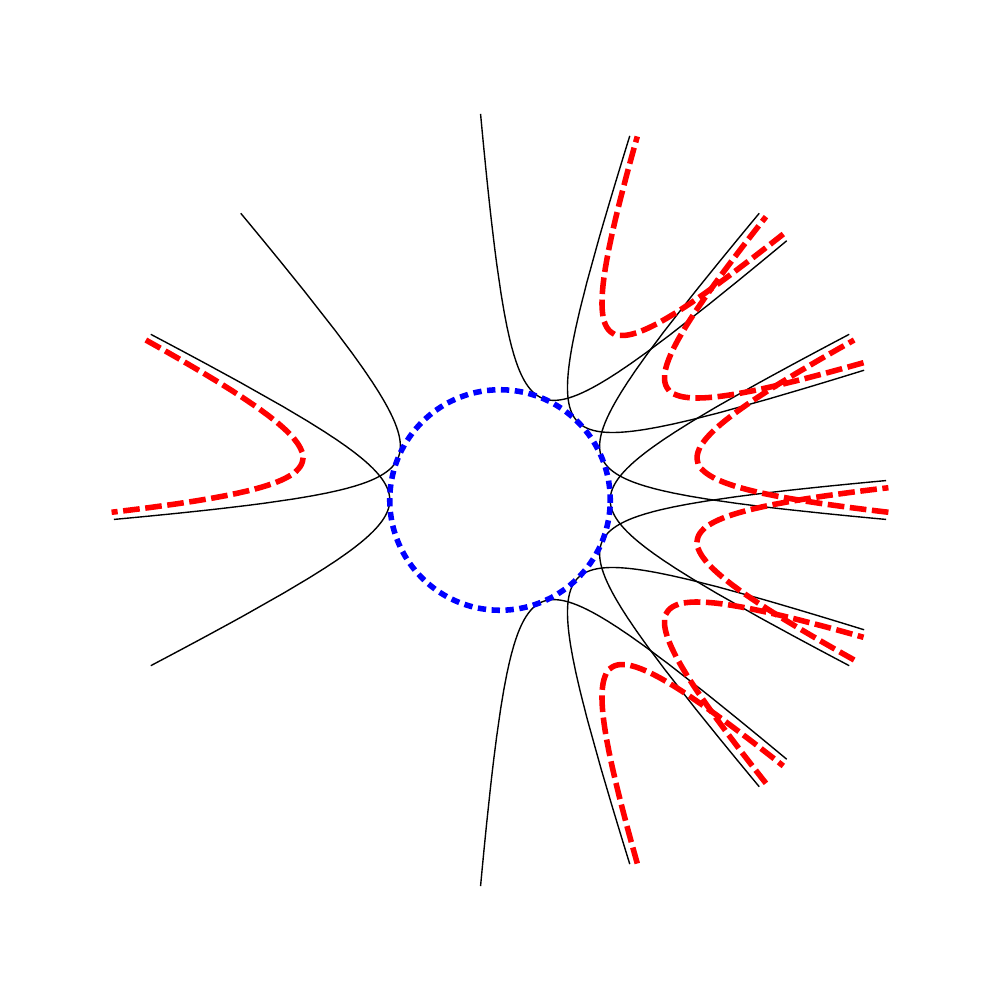} & 
\includegraphics[width=.31\textwidth]{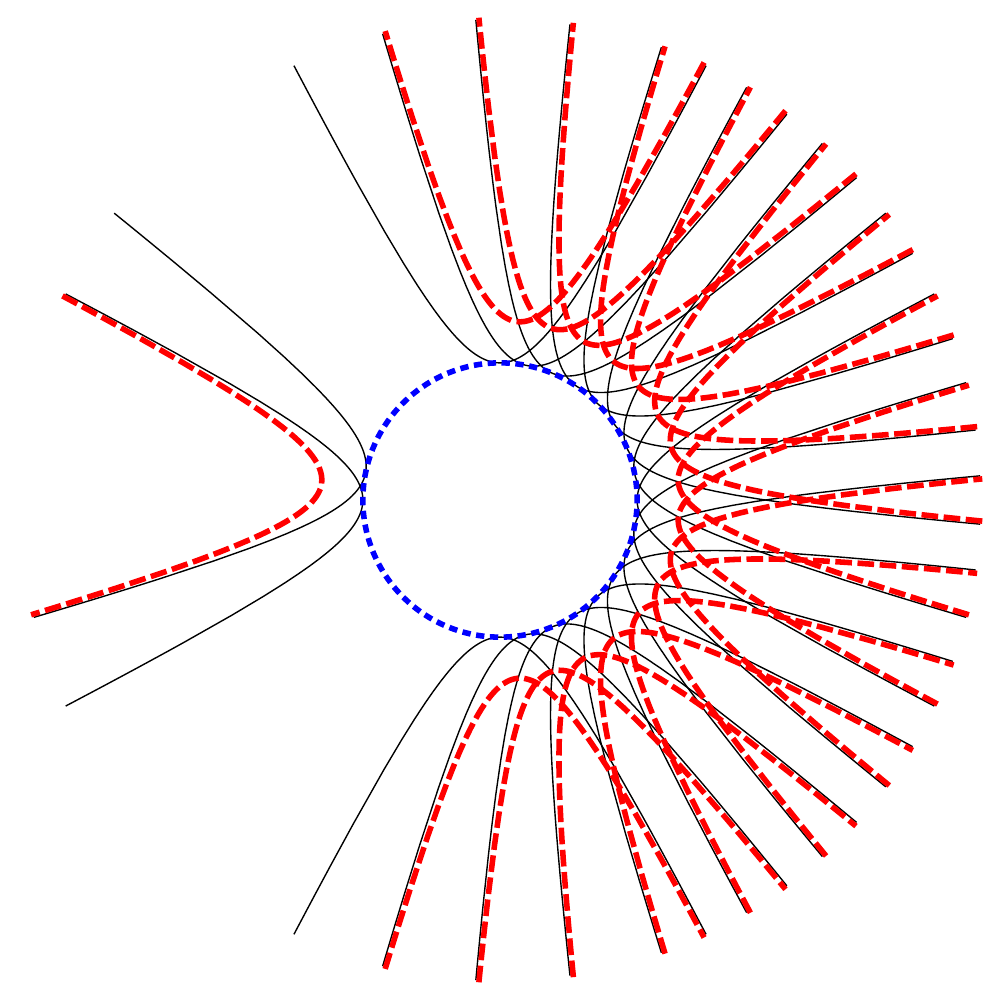} & 
\includegraphics[width=.31\textwidth]{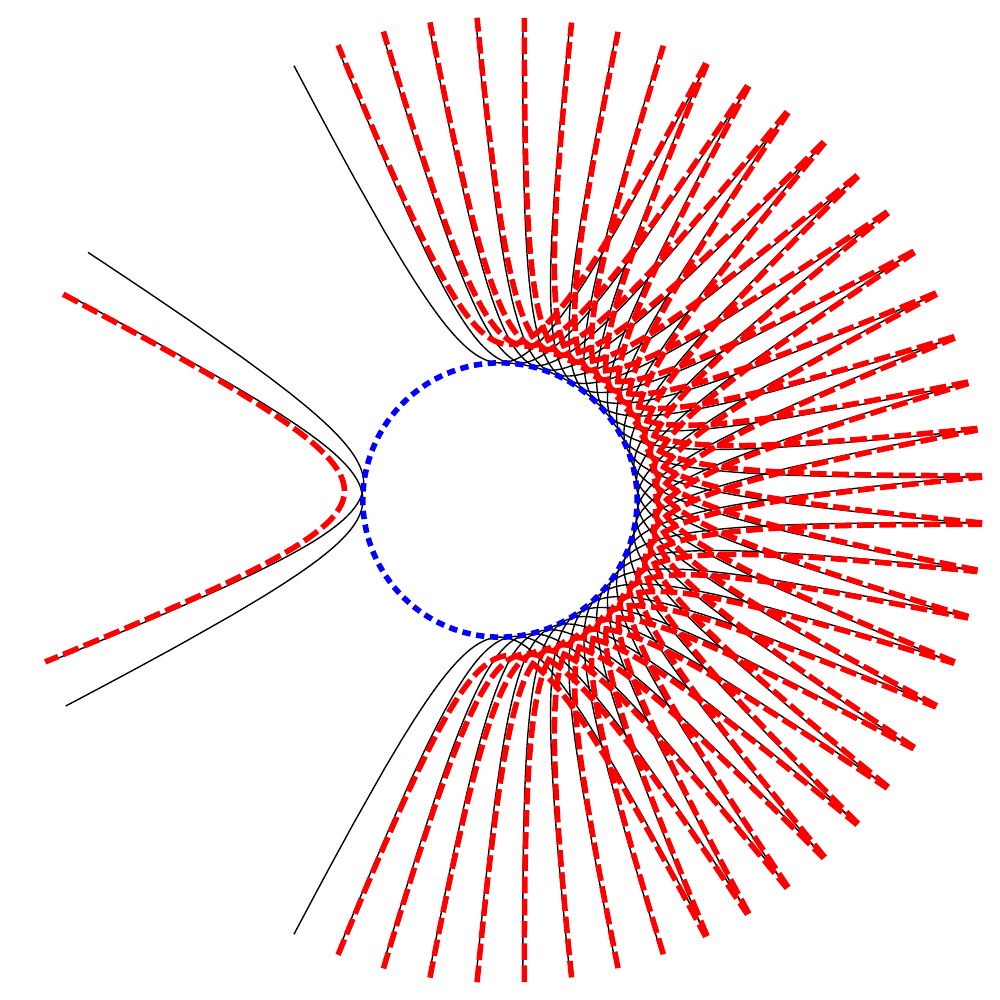}
\end{tabular}
\caption{The spatial geodesics that extend across intervals of length $2\alpha_0$ and $2\alpha_0-\pi/K$. In the limit $K \to \infty$, the differences between their lengths arise only from the tips of the geodesics and make up a circle of radius $R_0$ in the center. The graphs show $2K = 16, 32, 64$. For arbitrary curves the cancellations are more subtle (see below.)}
\label{boundpic}
\end{figure}

The result (\ref{minradius}) immediately provides a holographic explanation for the match between our proposed formula for differential entropy and the area of a hole in AdS$_3$.    Consider the discretized formula for $K$ observers (\ref{Kobservers}).    Pictorially, the entropy of the intervals of angular size $2\alpha_0$ is measured by the length of the black geodesics in Fig.~\ref{boundpic}.   Meanwhile, the subtraction terms are measured by the red geodesics in Fig.~\ref{boundpic}.    The difference in the lengths of the black and red curves comes entirely from the near-tip segments of the black geodesics.  In the limit that $K \to \infty$ these segements form a circle of radius $R_0$ and mark the boundary of the hole in the spherical Rindler-AdS space.

\paragraph{Differential entropy and the strong subadditivity bound:}
The argument above is reminiscent of the holographic proof of the strong subadditivity bound for quantum information \cite{headrick}.    This is not a coincidence.  In fact, our formula for differential entropy precisely saturates the strong subadditivity bound on the information in a union of short intervals.\footnote{The observation in this paragraph applies when $\alpha_0 < \pi / 4$.} In fact, this result holds even for non-holographic theories. To see this, recall that given spatial intervals $I_1$ and $I_2$, the entanglement entropy of the union $I_1 \cup I_2$ satisfies the bound
\begin{equation}
S({I_1 \cup I_2}) \leq S(I_1) + S(I_2) - S(I_1 \cap I_2) \, .
\label{strsub}
\end{equation}
Applied to intervals of angular size $2\alpha_0$ and separation $\Delta\theta$, the bound coincides with differential entropy of two observers (\ref{twoobserver}).  Iterating this formula for $K$ evenly spaced intervals on a line gives:
\be
S(\cup_{j=1}^{K} I_j) \leq 
\sum_{j=1}^{K} S(I_j) - \sum_{j=1}^{K-1} S(I_j \cap I_{j+1}) \, .
\label{subboundint}
\ee
An analogous quantity bounds the entanglement entropy of $\cup_{j=K+1}^{2K} I_j$. We now combine the two unions to form a circle. The overlap term consists of two disconnected pieces: $I_K \cap I_{K+1}$ and $I_{2K} \cap I_1$. When $\alpha_0 < \pi / 4$, the entanglement entropy of this bipartite overlap region is the sum of the entanglement entropies of the two parts. Applying strong subadditivity for the final time, we obtain:
\be
S(\cup_{j=1}^{2K} I_j) \leq 
\sum_{j=1}^{2K} S(I_j) - \sum_{j=1}^{2K} S(I_j \cap I_{j+1})  = \sum_{j=1}^{2K} \big(S(I_j) - S(I_j \cap I_{j+1})\big) \, ,
\label{subbound}
\ee
where $I_{2K+1} \equiv I_1$. For intervals of angular size $2\alpha_0$, the bound on the right hand side precisely reproduces our proposed formula for differential entropy (\ref{Kobservers}).   This is rather surprising, because  (\ref{subbound}) is trivial as a bound
on the entanglement entropy of the union of intervals that cover a Cauchy slice of a theory in a pure state, because in this case the left hand side of (\ref{subbound}) vanishes.    This suggests a novel interpretation of strong subaddivity: $E \geq S_{ent}$ where   $E$ is the differential entropy associated to local, finite-time observers and $S_{ent}$ is the entanglement entropy of their associated causal domains.  As a final corollary, note that $S_{ent}$ of a spatial region is equal to the differential entropy when the region of the field theory in which measurements are performed covers the whole domain of dependence of the spatial region.

\subsection{Differential entropy and the area of  arbitrary closed curves}
\label{irregular}

Our formula for differential entropy (\ref{contobs})  can be generalized to a family of observers whose time intervals vary continuously as a function of their spatial location.  A discrete version of this problem is defined by a collection of $2K$ evenly spaced finite segments $I_j$, whose sizes we denote $2\alpha_j$:
\begin{equation}
I_j = \left( \frac{\pi j}{K} - \alpha_j,\, \frac{\pi j}{K} + \alpha_j \right)  
\label{defiofj}
\end{equation}
Let us try formula~(\ref{subbound}), which is sufficiently general to apply to our current setup. We obtain:
\be
E \stackrel{?}{=}  
\sum_{j=1}^{2K} \big(S(I_j) - S(I_j \cap I_{j+1})\big) =  \frac{c}{3} \,\sum_{j=1}^{2K} \left( \log \frac{2L}{\mu} \sin\alpha_j - 
\log \frac{2L}{\mu} \sin \frac{\alpha_j + \alpha_{j+1} - \pi/K}{2} \right) \, .
\label{genbounddiscr}
\ee
To take the continuum limit, we replace $\pi / K$ with $d\theta$:
\begin{align}
E & = \,\,\frac{c}{3} \int_0^{2\pi} \left(
\log\sin\alpha(\theta)-\log\sin\frac{\alpha(\theta)+\alpha(\theta+d\theta)-d\theta}{2}\right) \nonumber \\
& = \,\,\frac{c}{6} \int_0^{2\pi} \left(
2\log\sin\alpha(\theta)-\log\sin\frac{\alpha(\theta)+\alpha(\theta+d\theta)-d\theta}{2}
-\log\sin\frac{\alpha(\theta-d\theta)+\alpha(\theta)-d\theta}{2}\right) \nonumber \\
&= \,\,\frac{c}{6} \int_0^{2\pi} d\theta\, \cot\alpha(\theta)
= \frac{1}{2} \int_0^{2\pi} d\theta \,{d S(\alpha) \over d\alpha}\Big{|}_{\alpha(\theta)}.
\label{genbound}
\end{align}

\begin{figure}[t!]
\centering
\begin{tabular}{cc}
\raisebox{10mm}{\includegraphics[width=.45\textwidth]{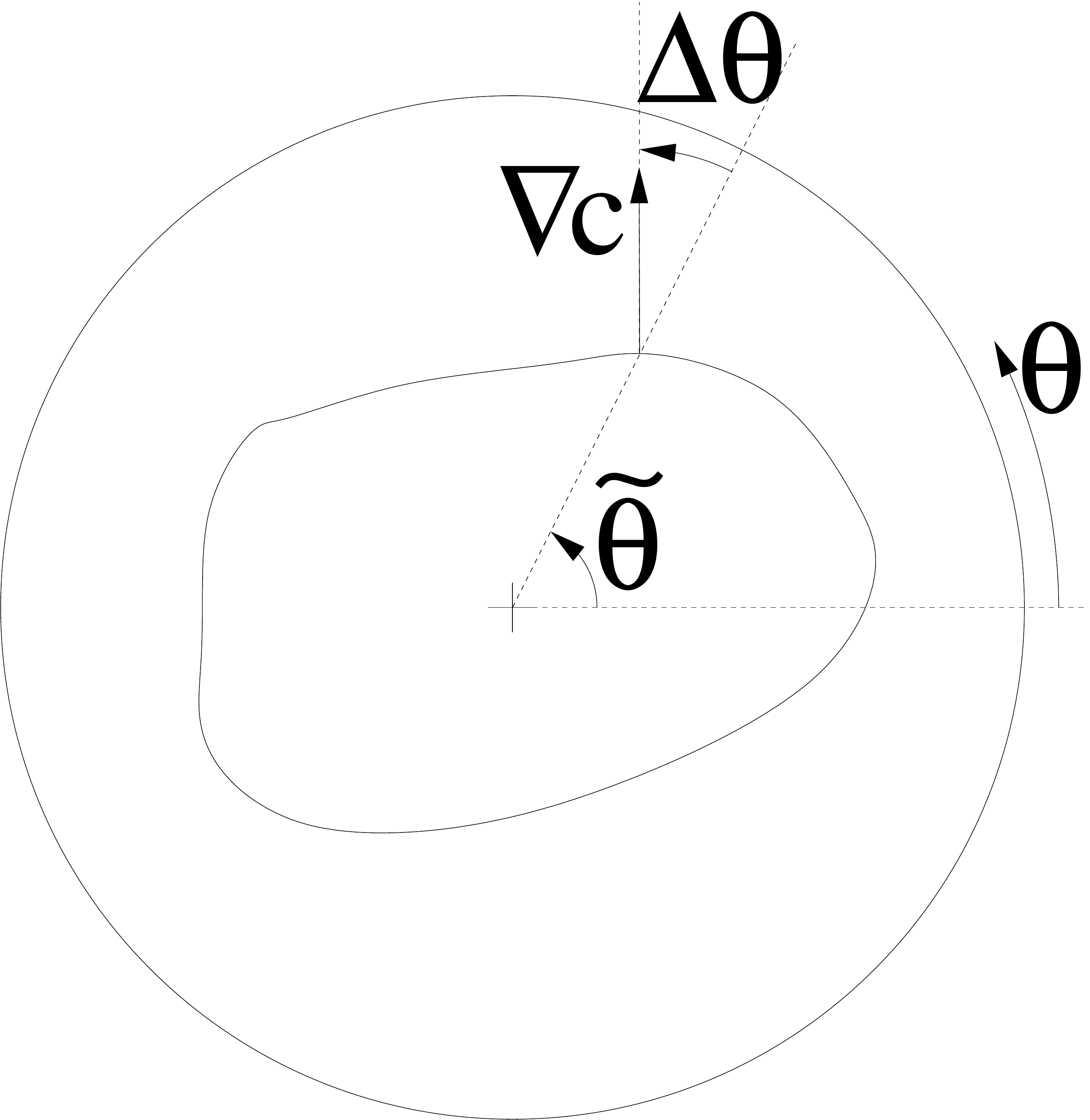}} & 
\includegraphics[width=.55\textwidth]{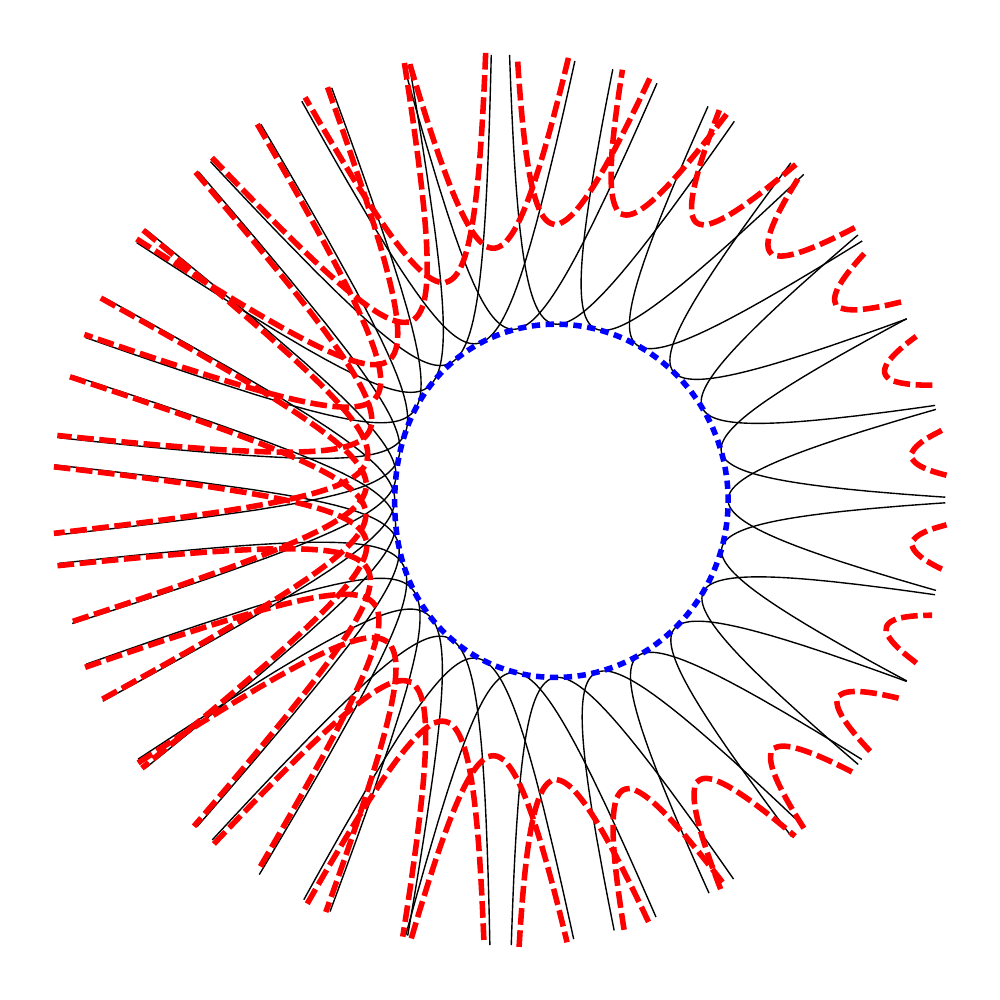}
\end{tabular}
\caption{Left: the notation of eqs.~(\ref{curve}-\ref{Deltatheta}). Right: the geodesics, which make up eq.~(\ref{genbounddiscr}).}
\label{thetas}
\end{figure}

Previously we found that the duration of observations in the field theory was related to the radius of a corresponding hole in spacetime.  Hence it should be the case that the differential entropy formula (\ref{genbound}) reproduces the area of an inhomogeneous hole in AdS$_3$.   To test this, consider a closed curve on a spatial slice of AdS$_3$:
\begin{equation}
c(R, \tilde\theta) = R - \tilde R(\tilde{\theta}) = 0.
\label{curve}
\end{equation}
We parametrize the curve by  $\tilde\theta$, reserving $\theta$ for the angular coordinate on the boundary field theory (see the left panel in Fig.~\ref{thetas}).   The length of the curve (\ref{curve}), which we denote $\mathcal{A}$, evaluates to
\begin{equation}
\frac{\mathcal{A}}{4G}= \frac{1}{4G} \int_0^{2\pi} d\tilde\theta \,
\sqrt{R^2 + \left(1 + \frac{R^2}{L^2} \right)^{-1} \left( \frac{dR}{d\tilde\theta}\right)^2}
= \frac{1}{4G} \int_0^{2\pi} \frac{d\tilde\theta\, R}{\cos\Delta\theta}.
\label{length}
\end{equation}
We wish to isolate a time strip of the field theory, in which local observers contain complete information about the exterior of $c(R,\tilde\theta) = 0$. Generalizing the discussion of Sec.~\ref{finitetime}, the exterior of the hole can be probed by a continuous family of Rindler observers whose acceleration horizons are tangent to the curve (\ref{curve}). The trajectories of these observers asymptote to the null rays projected orthogonally from the curve.

%\includegraphics[width=.55\textwidth]{inhomogeneous_hole.pdf}
%\end{tabular}
%\caption{Left: the notation of eqs.~(\ref{curve}-\ref{Deltatheta}). Right: the geodesics giving positive contribution to eq.~(\ref{lengthbound}) are tangent to the hole.}

At $\tilde\theta$, the vector orthogonal to (\ref{curve}) is:
\begin{equation}
\vec\nabla c = \left(1 + \frac{R^2}{L^2}\right)\frac{\partial}{\partial R} 
+ \frac{1}{R^2}\frac{\partial \tilde{R}}{\partial \tilde\theta} \frac{\partial}{\partial \tilde\theta}.
\end{equation}
The angle it makes with the radial vector $\partial / \partial R$ is given by:
\begin{equation}
\cos\Delta\theta(\tilde\theta) 
= \left(1 + \frac{L^2}{L^2 + \tilde{R}(\tilde\theta)^2}
\left( \frac{d\log\tilde{R}(\tilde\theta)}{d\tilde\theta}\right)^{\! 2}\, \right)^{-1/2}.
\label{Deltatheta}
\end{equation}
We take $\Delta\theta$ to have the opposite sign from $d \tilde{R}/d\tilde\theta$.
Eq.~(\ref{hitboundary}) from Appendix~\ref{nulls} tells us that a null ray projected in this relative direction from the point $\big(\tilde{R}(\tilde\theta), \tilde{\theta}\big)$ hits the boundary at
\begin{equation}
T(\tilde\theta) = L \cot^{-1}\frac{\tilde{R}(\tilde\theta) \cos\Delta\theta(\tilde\theta)}{L}  
\equiv L\,\alpha(\tilde{\theta})
\qquad {\rm and} \qquad
\theta(\tilde\theta) = \tilde\theta + 
\tan^{-1} \frac{L\tan\Delta\theta(\tilde\theta)}{\sqrt{L^2 + \tilde{R}(\tilde\theta)^2}}.
\end{equation}
From here on we write $R$ for $\tilde{R}(\tilde\theta)$. The size and location of the boundary causal diamond simplify to:
\begin{align}
\alpha(\tilde\theta) & = \tan^{-1} \frac{L}{R}\sqrt{1 + \frac{L^2}{L^2+R^2} \left(\frac{d\log R}{d\tilde\theta}\right)^2} \label{balpha} \\
\theta(\tilde\theta) & = \tilde\theta - \tan^{-1} \frac{L^2}{L^2+R^2} \frac{d\log R}{d\tilde\theta}
\label{btheta}
\end{align}
Alternative derivations of eqs. (\ref{balpha}) and (\ref{btheta}) can be found in Appendix~\ref{sec.altder}. Using these expressions in the formula for differential entropy (\ref{genbound}), we obtain:
\begin{equation}
E =  
\frac{L}{4G} \int_0^{2\pi} d\theta \cot\alpha(\theta)
= \frac{1}{4G} \int_0^{2\pi} d\tilde\theta \,
\frac{d\theta}{d\tilde\theta} \,
R\cos\Delta\theta.
\label{lengthbound}
\end{equation}
Eqs.~(\ref{lengthbound}) and (\ref{length}) do not look similar.

%As in the case of a spherical hole, this expression is a combination of lengths of spatial geodesics corresponding to appropriate boundary causal diamonds. This is illustrated in the right panel of Fig.~\ref{thetas}, in analogy to Fig.~\ref{boundpic}.

\subsection{Proof that differential entropy reproduces lengths of curves}

\begin{figure}[t!]
\centering
\begin{tabular}{ccccc}
\includegraphics[width=.25\textwidth]{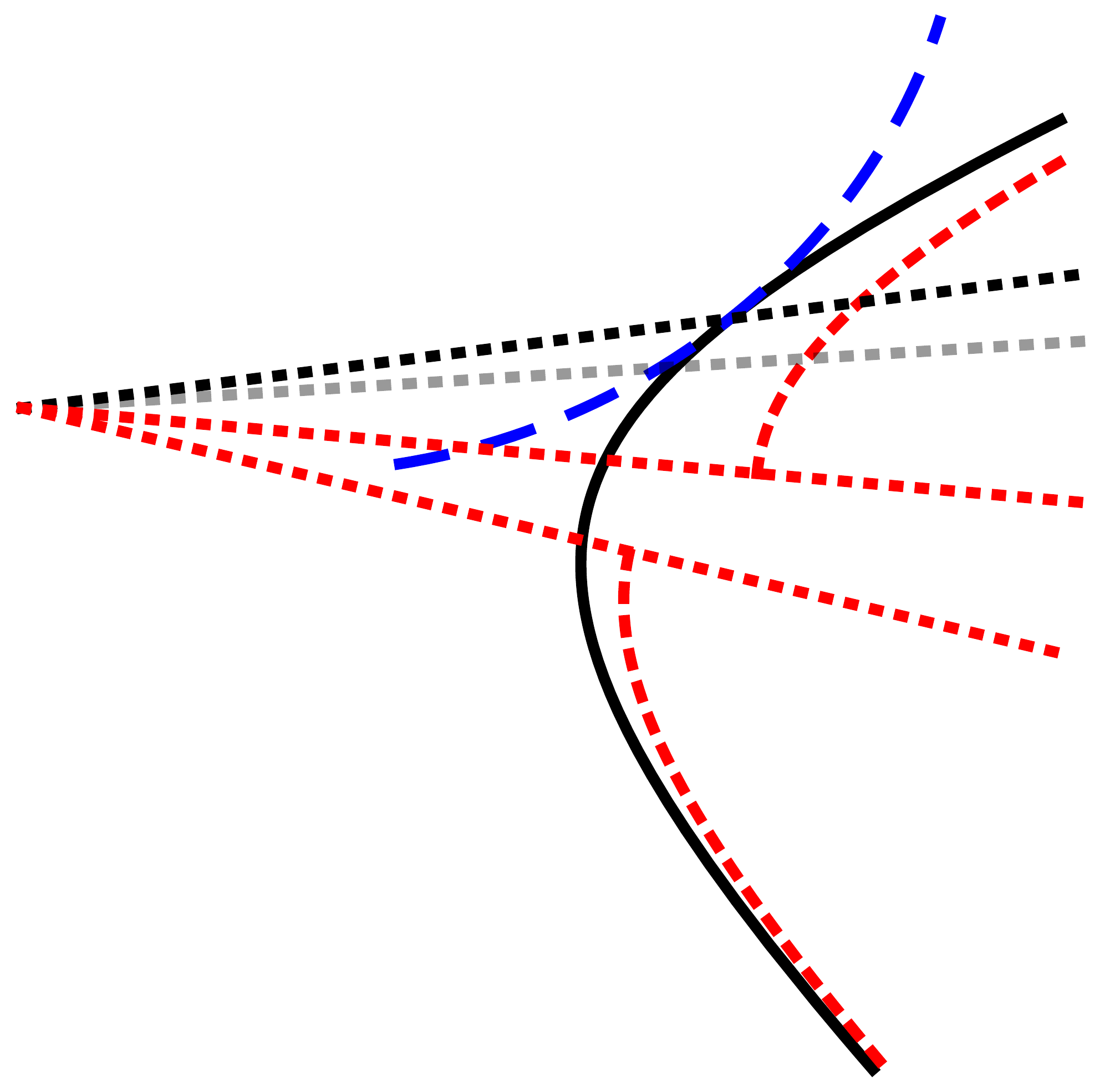} 
& \qquad \raisebox{2cm}{+} \qquad & 
\includegraphics[width=.25\textwidth]{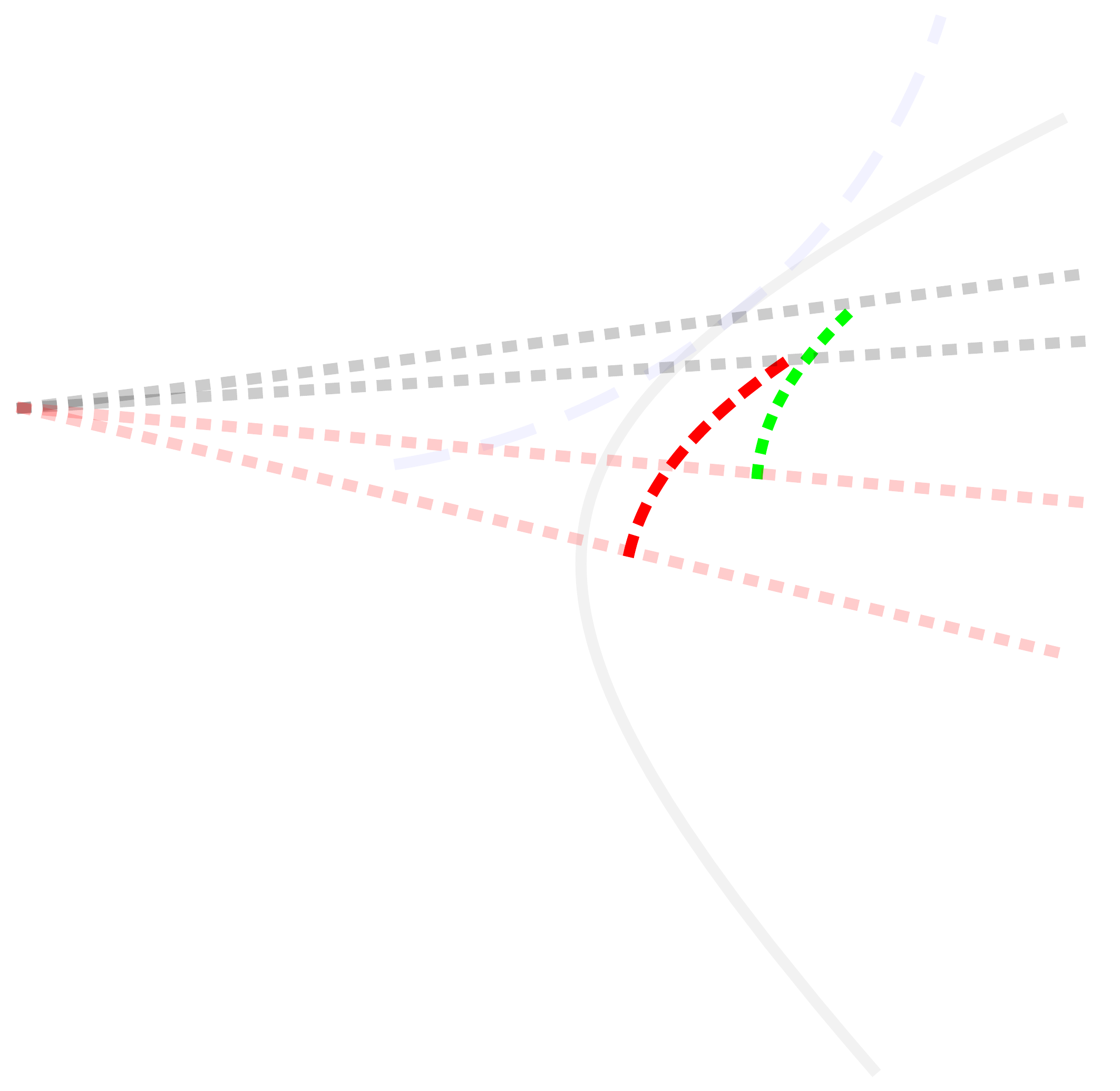} 
& \qquad \raisebox{2cm}{=} \qquad  & 
\includegraphics[width=.25\textwidth]{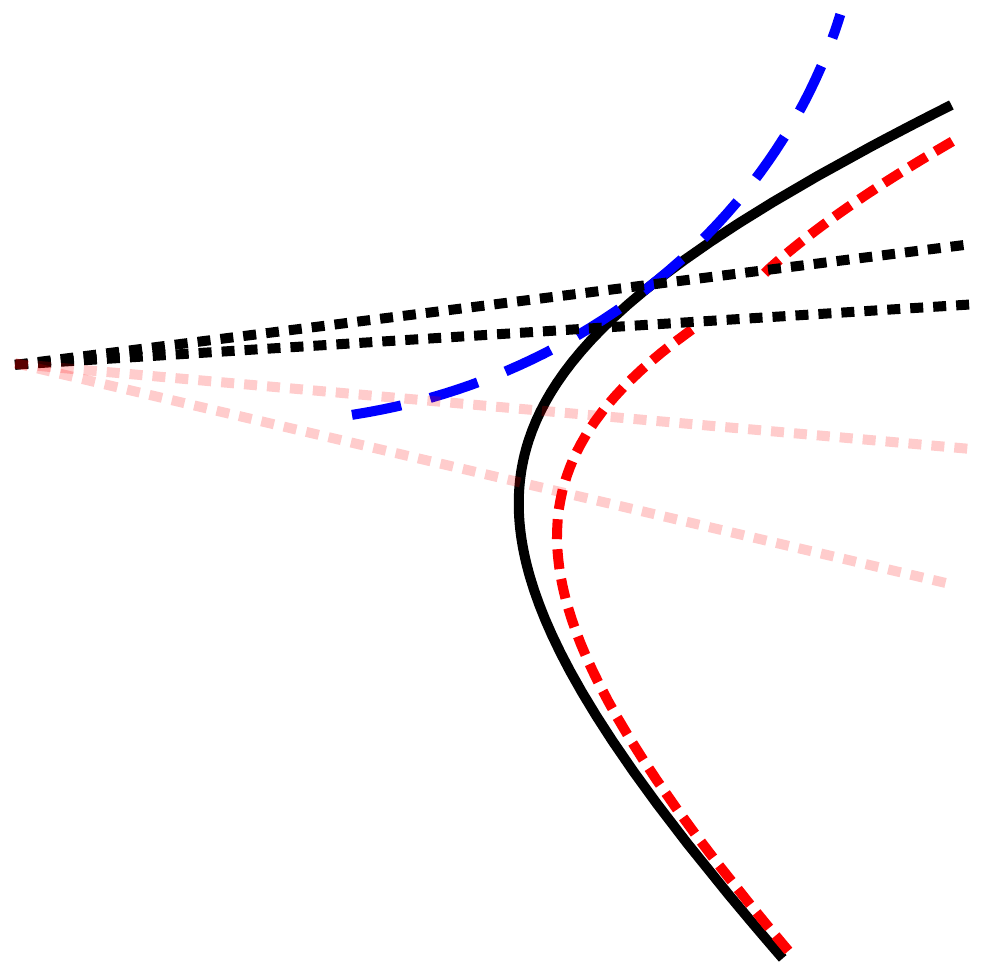}
\end{tabular}
\caption{Left: The integrand in eq.~(\ref{genbound}) is the length of the black, continuous geodesic minus the lengths of the red, dashed half-geodesics. The angle between the two straight red lines is $d\theta$. Center: The form (\ref{thecure}) adds the red, thickly dashed length and subtracts the green, finely dotted length. Right: The resulting integrand is that in eq.~(\ref{newintegrand}), the length element along the curve $R(\tilde\theta)$. The angle between the two straight black lines is $d\tilde\theta$.}
\label{proof}
\end{figure}

To prove the equivalence of eqs.~(\ref{length}) and (\ref{genbound}) (or eq.~\ref{lengthbound} after substitutions), add to (\ref{genbound}) the integral of the exact form:
\begin{equation}
\frac{c}{6} \cdot d \left(\frac{1}{2}\log\frac{\sin (\alpha-(\theta-\tilde\theta))}{\sin (\alpha+(\theta-\tilde\theta))}\right)
\label{thecure}
\end{equation}
The choice of form is geometrically motivated. The expression in the parentheses is the length of the spatial geodesic between the angular location $\theta(\tilde\theta)$ and $\tilde\theta$. Adding this form to the integrand reshuffles the negative contributions to (\ref{genbound}) without changing its total value. The geometric effect of adding (\ref{thecure}) is illustrated in Fig.~\ref{proof}. 

Adding the integral of (\ref{thecure}) to (\ref{genbound}), we obtain:
\begin{equation}
E = \frac{c}{6} \int_{\tilde\theta=0}^{\tilde\theta=2\pi} \,\frac
{\sin 2 (\theta-\tilde\theta) \,d\alpha -2 \cot\alpha \sin^2 (\theta-\tilde\theta)\, d\theta + \sin 2 \alpha \,d\tilde\theta}
{\cos 2(\theta-\tilde\theta)-\cos 2\alpha}
\label{newintegrand}
\end{equation}
After plugging in eqs.~(\ref{balpha}-\ref{btheta}) and $c=3L/2G$, this reproduces the length of the curve in AdS$_3$ given in (\ref{length}). Note that the agreement extends to nonconvex curves, which cannot define a (nonhomogeneous) ``spherical'' Rindler-AdS space of the type discussed in Sec.~\ref{coords}.

Our proof has an immediate corollary. Suppose we attempted to compute the length of an open curve in AdS$_3$ using formula~(\ref{lengthbound}). The mistake we would have made is:
\begin{align}
L \int_{\theta_i}^{\theta_f} d\theta\, & \cot\alpha(\theta) - \int_{\tilde\theta_i}^{\tilde\theta_f} d\tilde\theta \,
\sqrt{R^2 + \left(1 + \frac{R^2}{L^2} \right)^{-1} \left( \frac{dR}{d\tilde\theta}\right)^2} \\
& = L \int_{\tilde\theta_i}^{\tilde\theta_f}  
d \left(\frac{1}{2}\log\frac{\sin (\alpha+(\theta-\tilde\theta))}{\sin (\alpha-(\theta-\tilde\theta))}\right) = \frac{L}{2}
\log\frac{\sin (\alpha+(\theta-\tilde\theta))}{\sin (\alpha-(\theta-\tilde\theta))}
\,\Bigg|_{\tilde\theta_i}^{\tilde\theta_f}
\nonumber 
\end{align}
This vanishes if $\theta(\tilde\theta) = \tilde\theta$, which happens when:
\begin{equation}
\frac{d R(\tilde\theta_i)}{d\tilde\theta} = \frac{d R(\tilde\theta_f)}{d\tilde\theta} = 0
\label{opencond}
\end{equation}
We have learned that formula~(\ref{lengthbound}) applies also to open curves, which satisfy condition~(\ref{opencond}) at both endpoints. This opens the possibility of associating a differential entropy to field theory regions, which are bounded both in space and time.

\section{Discussion}
\label{discussion}

In this paper we have proposed a new notion of uncertainty called differential entropy, which applies to field theory regions that are bounded in time. It is a measure of the amount of information about the system, which is inaccessible to local observers.   The new notion raises many interesting issues and open questions, some of which we discuss below.

\paragraph{Interpreting differential entropy: } We set out to quantify the uncertainty about a state given the outcomes of a set of local measurements in field theory. In the quantum information theory literature, a formal solution of this problem was presented in \cite{donald}. 
%a related quantity applying to the case where the restricted measurements form a subalgebra of %the complete set, has been studied in \cite{subalgebras, cntrev}. 
In a holographic context, Kelly and Wall \cite{wall} have argued that causal holographic information \cite{CHI} of a spatial region in the boundary  quantifies the uncertainty left over after  local observers measure all one-point functions in the associated causal domain (see also \cite{2bens}).   This notion is clearly related to the concept that we have attempted to capture in this article.  Our focus has been on domains which are bounded in time, but are not necessarily causal domains of any spatial region. One hint we may take from the proposal of \cite{wall} is that differential entropy may be more directly related to causal holographic information than to entanglement entropy. In the settings of the present work -- the vacuum of AdS$_3$ -- these two concepts agree, but they differ in more general situations \cite{CHI}.

Another related concept is topological entanglement entropy \cite{topentr, topentr2}, which has been exploited in the holographic context in \cite{verlindes1, verlindes2, verlinde3}. Its definition mirrors our requirement that the differential entropy be UV-finite. This appears to be associated with the universal contribution to the entanglement entropy (see ``Going to higher dimensions'' below.)

\paragraph{Residual uncertainty and the time-energy uncertainty:} 
Our arguments have assumed that an accelerated observer has access to the full reduced density matrix of the associated boundary causal diamond. This is the maximal amount of information that the observer can access without violations of bulk causality. From the boundary point of view, a potentially more stringent restriction arises from the time-energy uncertainty relation. We expect that observers who make measurements over a time interval $\Delta T$ can resolve energy differences of order $(\Delta T)^{-1}$, but not smaller. Consider a boundary observer who makes measurements over a time interval of order $2 \pi L$, i.e. the minimal interval sufficient for observing a whole Cauchy slice. Such an observer should be able to distinguish the ground state from the first excited state ($\Delta E \sim L^{-1}$), but may not be able to distinguish one excited state from another -- because there energy splittings can be exponentially small. This qualitative argument is consistent with the result that the differential entropy vanishes when $\alpha_0 = \pi/2$. It also implies that in thermally excited states the differential entropy should not vanish until the time strip covers an exponentially long time -- an expectation, which is borne out by a simple calculation in the BTZ geometry.  This is also consistent with the argument in \cite{vijaydonmoshe} that information about the precise microstate of a black hole can always be recovered by sufficiently precise measurements of the mass from infinity, but that this requires exponentially long timescales that are inaccessible to conventional semiclassical observers, thus leading to effective entropy and information loss.

\paragraph{Differential entropy in BTZ: }  It is not obvious that formula (\ref{genbound}) will apply to excited states without modification. However, as a motivational example, consider a regular time strip in a thermal state, which should be dual to a circular hole in the BTZ spacetime:
\begin{equation}
ds^2 = -\frac{R^2 - R_+^2}{L^2} \,dT^2 + \frac{L^2}{R^2 - R_+^2} dR^2 + R^2 d\tilde\theta^2.
\label{btzmetric}
\end{equation}
A light ray projected orthogonally from an $R_0$-sized hole reaches the boundary at a time $T_0 = (L^2 / R_+) \coth^{-1} R_0/R_+$. Substituting the thermal state entanglement entropy
\begin{equation}
S(\alpha_0) = \frac{c}{3} \log \frac{2 L^2 \sinh (R_+\alpha_0/L)}{R_+ \mu}
\label{BTZgeodesiclength}
\end{equation}
for $\alpha_0 = T_0 / L$ in formula (\ref{genbound}) returns the area of the hole in appropriate units: $E = 2\pi R_0 / 4G$ (here again $c = 3L/2G$). The apparent success of this formula is puzzling, because it extends beyond the regime of validity of eq.~(\ref{BTZgeodesiclength}). In particular, we still recover the circumference of the hole when $\alpha_0$ exceeds the critical value at which the Araki-Lieb inequality is saturated (the regime dubbed the ``entanglement plateau'' in \cite{plateau}) and even when $\alpha_0 > 2\pi$ -- that is to say, when the relevant spatial geodesics no longer compute entanglement entropies. Note that arbitrarily small holes $R_0 \to^+ R_+$ translate to $\alpha_0 \gg 2\pi$ -- which resonates with the observations of the previous paragraph. In sum, it appears that the differential entropy of excited states likely involves a more general quantity than entanglement entropy, whose definition incorporates the constraints of the time-energy uncertainty relation. The correct generalization of (\ref{genbound}) may offer a holographic interpretation of the focusing theorem \cite{focusingthm}. The problem is related to identifying a holographic dual to a density matrix, which has been discussed in \cite{probingcurves, densitydual, bensold, properties} (see also \cite{recentBTZ}). A helpful intermediate step might be to covariantize the proposal (\ref{genbound}) using the time-dependent generalization of the Ryu-Takayanagi relation \cite{hrt}.

\paragraph{Going to higher dimensions: } 
An obvious direction for future work is to lift our calculation to higher dimensions. To appreciate why this is challenging, recall that our derivation of eq.~(\ref{genbound}) involved decomposing a finite time strip of a two-dimensional field theory into a union of causal diamonds. On the time slice of symmetry, this translates into decomposing a circle into a union of overlapping intervals. In higher-dimensional settings, one would have to cover a sphere with disks, but an overlap of two disks does not have a regular shape. The appearance of ``corners'' reflects a richer structure of divergences in the entanglement entropy in higher dimensions \cite{liumezei}, which are more difficult to cancel out than the simple logarithmic divergence of eq.~(\ref{oneobserver}). Conversely, a generalization of formula (\ref{genbound}) to more dimensions is likely to carry a deep lesson about how the emergence of space ties in with the holographic renormalization group. For two-dimensional field theories, one such lesson is offered by \cite{Casini:2006es} and \cite{headrick}: the first of these papers relates the $c$-theorem to strong subadditivity while the second relates strong subadditivity to general properties of geodesics in the bulk spacetime.
Some relevant references include \cite{Myers:2010tj, Myers:2012ed, Casini:2012ei, Klebanov:2012yf}; see also \cite{nima} on the holographic emergence of Einstein's equations.

\paragraph{Effective Hamiltonian for differential entropy: }  We have defined differential entropy in terms of the finite-time measurements accessible to local observers.   There may be many underlying pure or mixed states that give rise to a particular set of such measurements.  It would be good to characterize this set of states in generality.  One special state in this class is a mixed state with maximal von Neumann entropy.\footnote{We thank Matt Headrick for a discussion on this point.} The density matrix of this distinguished state can be written as \cite{wall, hartle}
\be
\rho = Z^{-1} e^{-\sum_m \lambda_m \mathcal{O}_m} .
\label{effham}
\ee
The $\mathcal{O}_m$ are observables whose expectation values are given and the $\lambda_m$ are a set of chemical potentials, which are selected by these expectation values.    The exponent on the right hand side has the structure of an effective Hamiltonian.\footnote{If $\rho$ were a reduced density matrix of some tensor factor of the Hilbert space, this would be the modular Hamiltonian.}   In our case $\mathcal{O}_m$ are observations made by local observers over a finite time $T_0$, which are hence consistent with interactions that span only a finite spatial range $\alpha_0$.   Thus, the effective Hamiltonian defined by the exponent of (\ref{effham}) might be non-local, but will contain interactions that have a maximal spatial  range $\alpha_0$.  Holographically, the idea is that the interior of the hole is related to the highly non-local interactions in this Hamiltonian (see \cite{albionmonica} for related comments in the setting of the holographic renormalization group).  It appears that the differential entropy could be computed by writing down the most general Hamiltonian $H$, which contains interactions over the same distance as those over which observers can make measurements. If we insist that  the reduced density matrices obtained from $\rho=e^{-H}$ (after normalization) are exactly the ones that were a priori given, this uniquely fixes $H$. Finally, the entropy is simply the entropy associated
to $\rho=e^{-H}$, i.e. that of a system described by $H$ at finite temperature! We notice a suggestive similarity with the structure of the reduced density matrix for Rindler space. One difference is that in the latter case the effective Hamiltonian turns out to be local \cite{casinihuertamyers}. This may be related to the fact that AdS-Rindler space is associated to a boundary causal diamond, which is not capped at any maximal time except as determined by causality.

The Hamiltonian that we obtained above depends on $\alpha_0$, the size of the spatial interval. Its interactions become longer range as we increase $\alpha_0$. This suggests that we should be able to interpret the change in $\alpha_0$ as some kind of RG flow. What type of  flow could this be? Local finite-time observers have access to short distance physics, but not to long distance physics. Therefore, by restricting to such observers we have effectively integrated out IR degrees of freedom with energies less than $1/\alpha_0$.  As we decrease $\alpha_0$, we are integrating out more and more IR degrees of freedom, so this  flow is exactly the opposite of standard  RG flow. The differential entropy then has a natural interpretation as a generalization of entanglement entropy between the UV and IR degrees of freedom.  Some types of UV-IR entanglement were studied previously in \cite{uvir}.  However, in our case the Hamiltonian $H$ and the density matrix $\rho=e^{-H}$ still act on the full Hilbert space of the CFT and not only on a tensor factor, and therefore there is not a precise interpretation as entanglement entropy between UV and IR degrees of freedom.   This also suggests that there may not be an effective field theory associated with low-energy gravity, where the Hilbert space factorizes between the interior and exterior of a closed surface.  From the viewpoint of Sec.~\ref{coords}, this follows directly from noting that unlike in the usual Rindler decomposition of Minkowski or AdS space, there is no set of coordinates, which would allow us to insert an object inside a hole without leaving an imprint on the outside. If one could quantify the departure of the Hilbert space from a tensor product ansatz, it would provide a systematic way to understand the breakdown of bulk low energy effective field theory and potentially resolve the recently publicized ``firewall'' problem of black holes \cite{Braunstein:2009my, firewalls}. It would be interesting to understand the factorization of the Hilbert space in the light of how local bulk observables are constructed on the boundary \cite{bdhm, kll}.

\paragraph{Related questions: } Although we phrased everything in terms of field theory and AdS/CFT, similar notions of differential entropy can be easily defined for all kinds of quantum mechanical systems such as spin chains, where they might be studied in much more detail. For a periodic spin chain, one could for example ask how much differential entropy the system has if one knows the reduced density matrices for all sequences of $L$ consecutive spins in the spin chain. This is an interesting new probe of such quantum mechanical systems, which to our knowledge has not been studied so far. 

These types of questions are closely related to a question in quantum information theory known as the quantum marginal problem \cite{margsummary}. There, one typically asks whether a state or a density matrix exists in case one is given reduced density matrices for various subsystems. For example, for a spin chain with $L=1$ wherein one only knows the reduced density matrices for the individual spins, \cite{klyachko} gave explicit necessary and sufficient conditions in terms of the eigenvalues of the reduced density matrices for the existence of a single pure state with the required projections. It would be interesting to study whether this  technology can be of use for the study of differential entropy as well.

%\acknowledgments
\paragraph{Acknowledgments: }
We thank Dionysios Anninos, Matthias Christandl, Ben Freivogel, Patrick Hayden, Matthew Headrick, Veronika Hubeny, Nima Lashkari, Stefan Leichenauer, Juan Maldacena, Don Marolf, Benjamin Mosk, Rob Myers, Maulik Parikh, Andrea Prudenziati, Eliezer Rabinovici, Mukund Rangamani, Simon Ross, Alejandro Satz, Masaki Shigemori, Ari Turner, Mark Van Raamsdonk, Erik Verlinde, and Herman Verlinde for helpful discussions. Part of this work was completed at the ``Complementarity, Fuzz, or Fire?'' rapid response workshop at KITP and at the ``Black Holes in String Theory'' workshop at the University of Michigan. VB was supported by DOE grant DE-FG02-05ER-41367 and by the Fondation Pierre-Gilles de Gennes. The work of BDC has been supported in part by the ERC Advanced Grant 268088-EMERGRAV. The work of BC has been supported in part by the Stanford Institute for Theoretical Physics. MPH acknowledges support from the Netherlands Organization for
Scientific Research under the NWO Veni scheme (UvA) and from the
National Science Centre under the grant 2012/07/B/ST2/03794 (NCNR). This work is part of the research programme of the Foundation for Fundamental Research on Matter (FOM), which is part of the Netherlands Organisation for Scientific Research (NWO).   

\appendix

\section{Radially accelerated trajectories}
\label{radials}
We wish to find radially accelerated trajectories in the metric (\ref{adsglobal}). Expressing the trajectory as $\xi^\mu(\tau)$ in terms of the proper time $\tau$, the equations that normalize the velocity and set the acceleration read
\begin{eqnarray}
-1 & = & g_{\mu\nu} \frac{d\xi^\mu(\tau)}{d\tau} \frac{d\xi^\nu(\tau)}{d\tau}\\
a^2 & = & g_{\mu\nu} \left(\frac{D}{D\tau} \frac{d\xi^\mu(\tau)}{d\tau}\right)  \left(\frac{D}{D\tau} \frac{d\xi^\nu(\tau)}{d\tau}\right).
\end{eqnarray}
For radially accelerated trajectories, $\xi^\mu$ varies over $T$ and $R$ and the conditions become:
\begin{align}
-1 & =
- \left( 1 + \frac{R(\tau)^2}{L^2}\right) \left(\frac{dT(\tau)}{d\tau}\right)^2 + 
\left( 1 + \frac{R(\tau)^2}{L^2}\right)^{-1} \left(\frac{dR(\tau)}{d\tau}\right)^2 
\label{normalization} \\
a^2 & = 
- \left( 1 + \frac{R(\tau)^2}{L^2}\right) 
\left( \frac{d^2T(\tau)}{d\tau^2} + \frac{2R(\tau)}{L^2 + R(\tau)^2} 
\frac{dT(\tau)}{d\tau}\frac{dR(\tau)}{d\tau}\right)^2 
\label{acceleration} \\
+ & \left( 1 + \frac{R(\tau)^2}{L^2}\right)^{-1}
\left( \frac{d^2R(\tau)}{d\tau^2} 
- \frac{R(\tau)}{L^2 + R(\tau)^2} \left(\frac{dR(\tau)}{d\tau}\right)^2
+ \frac{R(\tau)(L^2 + R(\tau)^2)}{L^4} \left(\frac{dT(\tau)}{d\tau}\right)^2
\right)^2
\nonumber
\end{align}
Eliminating $dT(\tau)/d\tau$ from (\ref{normalization}) and substituting in (\ref{acceleration}) gives
\begin{equation}
\frac{( R(\tau) + L^2 R''(\tau))^2}{R(\tau)^2 + L^2 \big(1 + R'(\tau)^2 \big)} = a^2 L^2.
\end{equation}
The simplest solution is
\begin{equation}
R(\tau) = R \qquad {\rm and} \qquad T(\tau) = \tau \left(1 + \frac{R^2}{L^2}\right)^{-1/2}\,,
\label{simplest}
\end{equation}
for which
\begin{equation}
a^2 L^2 = \frac{R^2}{L^2 + R^2} \label{accvsr}
\end{equation}
and $aL < 1$. We can relate this trajectory to all other radially accelerated trajectories with $aL<1$ by going to the embedding coordinates in $\mathbb{R}^{2,d}$
\begin{equation}
\left(\mathbf{T}_1, \mathbf{T}_2, \mathbf{R}\right) = 
(\sqrt{L^2 + R^2}\, \cos T/L, \sqrt{L^2 + R^2} \,\sin T/L, R)
\label{embed}
\end{equation}
and performing a boost in the $\mathbf{T}_1, \mathbf{R}$ variables
\begin{equation}
\mathbf{T}_1' = \frac{\mathbf{T}_1 + \mathbf{R} \cos\phi}{\sin\phi}
\qquad {\rm and} \qquad
\mathbf{R}' = \frac{\mathbf{R} + \mathbf{T}_1 \cos\phi}{\sin\phi}.
\end{equation}
The boost, which preserves the AdS hyperboloid \mbox{$\mathbf{T}_1^2 + \mathbf{T}_2^2 - \mathbf{R}^2 = L^2$}, defines new coordinates $T', R'$ in the same way as eq.~(\ref{embed}). Their relation to the old coordinates $T, R$ is
\begin{align}
R' & = \frac{R + \cos\phi \cos (T/L) \sqrt{L^2 + R^2}}{\sin\phi}, \\
T' & = L \cot^{-1} \left( \frac{\cot (T/L)}{\sin\phi} + \frac{R \cot\phi}{\sin (T/L) \sqrt{L^2 + R^2}} \right).
\end{align}
Under this transformation, the trajectory (\ref{simplest}) becomes
\begin{align}
R'(\tau) & = \frac{L}{\sin\phi\,\sqrt{1-a^2L^2}} 
\left(aL+\cos\phi \cos \frac{\tau\sqrt{1 - a^2 L^2}}{L}\right),
\label{rlessthan1} \\
T'(\tau) & = L \cot^{-1} \frac{aL\cos\phi+\cos\left((\tau/L) \sqrt{1-a^2 L^2}\right)}{\sin\phi \,\sin\left((\tau/L) \sqrt{1-a^2 L^2}\right)}.
\label{tlessthan1}
\end{align}
Here we have used eq.~(\ref{accvsr}) to eliminate $R$ and $T$ in favor of $a$. This is the general form of radially accelerated trajectories with $aL < 1$. These trajectories do not escape to the asymptotic boundary.

The trajectories with $aL > 1$ are obtained from eqs.~(\ref{rlessthan1}-\ref{tlessthan1}) by an analytic continuation. Denote $aL = \cos\rho$ and substitute $\rho \to i\rho$ and $\phi \to i\phi$. Dropping the primes and overall minus signs, the trajectories become:
\begin{align}
R(\tau) & = \frac{L}{\sinh\phi\,\sinh\rho} 
\left(\cosh\rho+\cosh\phi \cosh \frac{\tau\sinh\rho}{L}\right)
\label{defRtau} \\
T(\tau) & = L \cot^{-1} \frac{\cosh\phi\cosh\rho+\cosh\left((\tau/L) \sinh\rho\right)}{\sinh\phi \,\sinh\left((\tau/L) \sinh\rho\right)}
\label{defTtau}
\end{align}
For setting up the spherical Rindler coordinate system, it is convenient to parameterize the trajectories with
\begin{equation}
t = \frac{\tau \sinh\rho}{L}.
\end{equation}
The trajectories (\ref{defRtau}-\ref{defTtau}) expressed in terms of $t$ are given in eqs.~(\ref{defR}-\ref{defT}) in the text.

\section{Null geodesics}
\label{nulls}
Here we derive the form of arbitrary null geodesics in anti-de Sitter space. It is convenient to start with the null geodesics passing through the origin
\begin{equation}
T = L \tan^{-1} \frac{R}{L} \qquad {\rm and} \qquad \theta = \theta_0.
\label{nullradialorigin}
\end{equation}
Since AdS is homogeneous, any null geodesic that passes through $(T, R, \theta) = (0, R_0, 0)$ can be mapped to (\ref{nullradialorigin}) with a change of coordinates that puts $(0, R_0, 0)$ at the origin. In the embedding coordinates
\begin{equation}
\left(\mathbf{T}_1, \mathbf{T}_2, \mathbf{X}_1, \mathbf{X}_2\right) = 
(\sqrt{L^2 + R^2}\, \cos T/L, \sqrt{L^2 + R^2} \,\sin T/L, R \cos\theta, R \sin \theta)\,,
\label{embed2}
\end{equation}
this is accomplished with a boost in the $\mathbf{T}_1, \mathbf{X}_1$ variables: \begin{equation}
\mathbf{T}_1' = \frac{\mathbf{T}_1 \sqrt{L^2 + R_0^2} - \mathbf{X}_1 R_0}{L}
\qquad {\rm and} \qquad
\mathbf{X}_1' = \frac{\mathbf{X}_1 \sqrt{L^2 + R_0^2} - \mathbf{T}_1 R_0}{L}.
\end{equation}
Relating the boosted embedding coordinates to $T', R', \theta'$ defined as in eq.~(\ref{embed2}), we obtain the requisite coordinate transformation:
\begin{align}
T' & = L \tan^{-1} 
\frac{L \sqrt{L^2 + R^2} \,\sin T/L}{\sqrt{(L^2 + R^2)(L^2 + R_0^2)}\, \cos T/L - R_0 R \cos\theta} \label{deftt} \\
\theta' & = \tan^{-1}
\frac{L R \sin\theta}{R \sqrt{L^2 + R_0^2}\, \cos\theta - R_0 \sqrt{L^2 + R^2}\, \cos T/L}.
\label{deftheta}
\end{align}
We shall not need the explicit form of $R'$, only the fact that the boundaries $R' \to \infty$ and $R \to \infty$ coincide. The inverse transformation, which maps the origin to $(0, R_0, 0)$, is of the same form as eqs.~(\ref{deftt}-\ref{deftheta}) except $R_0 \to -R_0$.

As a final step, we substitute the null geodesic (\ref{nullradialorigin}) into the inverse of eqs.~(\ref{deftt}-\ref{deftheta}):
\begin{align}
T'(R) & = L \tan^{-1} 
\frac{LR}{L \sqrt{L^2 + R_0^2}+ R_0 R \cos\theta_0} \\
\theta'(R) & = \tan^{-1}
\frac{L R \sin\theta_0}{R \sqrt{L^2 + R_0^2}\, \cos\theta_0 + L R_0}.
\end{align}
This null geodesic hits the boundary at:
\begin{equation}
T'(\infty) = L \cot^{-1} \frac{R_0 \cos\theta_0}{L}
\qquad {\rm and} \qquad
\theta'(\infty) = \tan^{-1} \frac{L}{\sqrt{L^2 + R_0^2}} \tan\theta_0.
\label{hitboundary}
\end{equation}

\section{Alternative derivations of the irregular time strip}
\label{sec.altder}
\paragraph{Alternative derivation}
An elegant check on eqs.~(\ref{balpha}-\ref{btheta}) is as follows. A Rindler observer whose acceleration horizon is tangent to the curve (\ref{curve}) at $\tilde\theta$ defines a boundary causal diamond centered at $\theta(\tilde\theta)$, which extends between $T = \pm L\alpha(\tilde\theta)$. This means that the geodesic distance between $(T, R, \theta)=\big(0, R(\phi), \phi\big)$ and $\big(L\alpha(\tilde\theta), \infty, \theta(\tilde\theta)\big)$ must be null for $\phi = \tilde\theta$ and spacelike in a $\phi$-neighborhood. Two events in metric (\ref{adsglobal}) are spacelike separated if
\begin{equation}
\cos{\frac{T_1 - T_2}{L}} -
\frac{L^2 + R_1 R_2 \cos{(\theta_1 - \theta_2)}}{\sqrt{(L^2 + R_1^2)(L^2 + R_2^2)}} > 0\,,
\label{criterion}
\end{equation}
with equality holding for null separated events. Substituting $\big(0, R(\phi), \phi)$ and $\big(L\alpha(\tilde\theta), \infty, \theta(\tilde\theta)\big)$ into this condition, we conclude that 
\begin{equation}
d_{\tilde\theta}(\phi) = \cos\alpha(\tilde\theta) - \frac{R(\phi) \cos{(\theta(\tilde\theta) - \phi)}}{\sqrt{L^2 + R(\phi)^2}}
\end{equation}
must attain the minimum value 0 at $\phi = \tilde\theta$. Eqs.~(\ref{balpha}-\ref{btheta}) solve this extremization problem.

\paragraph{The simplest derivation}
If we hold the left hand sides of eqs.~(\ref{balpha}-\ref{btheta}) fixed and vary $\tilde\theta$, we obtain a curve $R(\tilde\theta)$. This curve is the spatial geodesic (\ref{rtcurve}) re-centered at $\theta(\tilde\theta)$. This means that when two points on the bulk curve $c$ are tangent to a common spatial geodesic, they define the same boundary causal diamond. This is a consequence of the fact \cite{CHI} that in three bulk dimensions, the causal holographic information and the entanglement entropy for any boundary causal diamond agree. Using this fact, an alternative way to derive eqs.~(\ref{balpha}-\ref{btheta}) is to start with the re-centered spatial geodesic (\ref{rtcurve}) and its $\tilde\theta$-derivative
\begin{align}
\tan^2 (\theta - \tilde\theta) = \frac{R^2 \tan^2\alpha - L^2}{R^2+L^2} 
\qquad \Rightarrow \qquad \qquad
R^2 & = \frac{L^2 \sec^2(\theta - \tilde\theta)}{\tan^2\alpha - \tan^2 (\theta - \tilde\theta)}
\\
\frac{d \log R}{d\tilde\theta} & = \frac{\sin(\theta-\tilde\theta) \cos(\theta - \tilde\theta)}{\cos^2(\theta-\tilde\theta) - \cos^2 \alpha}
\end{align}
and solve for $\alpha$ and $\theta-\tilde\theta$ in terms of $R$ and $d\log R/d\tilde\theta$. Eqs.~(\ref{balpha}-\ref{btheta}) solve this system of equations.

\bibliographystyle{utphys}

\end{document}